\documentclass[11pt,a4paper]{article}

\usepackage{jheppub}
\usepackage[utf8]{inputenc}
\usepackage{subfig}

\begin{document}

\title{Anomalous transport and holographic momentum relaxation} 
\author[1]{Christian Copetti$^\dagger$ \note{email: christian.copetti@uam.es}} 
\author[2]{, Jorge Fernández-Pendás$^\dagger$ \note{email:  j.fernandez.pendas@csic.es}} 
\author[3]{, Karl Landsteiner$^\dagger$ \note{email: karl.landsteiner@uam.es}}
\author[4]{, Eugenio Meg\'{\i}as$^\ddagger$ \note{email: eugenio.megias@ehu.eus}} 
\affiliation{$^\dagger$Instituto de
      Física Teórica UAM/CSIC,\\c/ Nicolás Cabrera 13-15, Cantoblanco,
      28049 Madrid, Spain\\ $^\ddagger$Departamento de Física Teórica,
      Universidad del País Vasco UPV/EHU, Apartado 644, 48080 Bilbao,
      Spain} \date{Dated: \today} 
      
      \abstract{The chiral magnetic and
      vortical effects denote the generation of dissipationless
      currents due to magnetic fields or rotation. They can be studied
      in holographic models with Chern-Simons couplings dual to
      anomalies in field theory.  We study a holographic model with
      translation symmetry breaking based on linear massless scalar
      field backgrounds. We compute the electric DC conductivity and
      find that it can vanish for certain values of the translation
      symmetry breaking couplings.  Then we compute the chiral magnetic
      and chiral vortical conductivities. They are completely
      independent of the holographic disorder couplings and take the
      usual values in terms of chemical potential and temperature. To
      arrive at this result we suggest a new definition of energy-momentum 
      tensor in presence of the gravitational Chern-Simons
      coupling.}  

\maketitle
\flushbottom
\section{Introduction}\label{sec:intro}
At finite temperature and chemical potential anomalies give rise to new dissipationless transport phenomena, the chiral magnetic (CME) and chiral vortical effects (CVE) (see \cite{Kharzeev:2013ffa,Landsteiner:2016led} for recent reviews). 
Holography has played a major role to arrive at the modern understanding of these effects \cite{Newman:2005hd,Banerjee:2008th,Erdmenger:2008rm,Landsteiner:2011iq}. 

In holography the dynamics of strongly coupled quantum systems is mapped onto the gravitational dynamics
of higher dimensional black holes \cite{Ammon:2015wua, Zaanen:2015oix}. 
In its most common form the dynamics of the black hole horizon describes a fluid and this horizon fluid can be mapped to
the hydrodynamics of the dual strongly coupled quantum system. 
In the hydrodynamic setup momentum is an exactly conserved quantity. 
Momentum conservation gives rise to convective transport of charge and leads to infinite DC conductivities.
Having applications of holography to strongly coupled
condensed matter systems in mind, one wants however a setup in which momentum is not conserved 
resulting in finite DC conductivities.
This means that translation symmetry must be broken. A way of doing this in holography is 
by giving the graviton a mass \cite{Vegh:2013sk, Davison:2013jba, Blake:2013bqa, Blake:2013owa, Donos:2013eha}. 
A particular simple way of introducing such a mass is based on massless scalar fields
in AdS space with spatially linear profiles \cite{Andrade:2013gsa}. 

Such models have been analyzed in \cite{Grozdanov:2015qia} for theories dual to (2+1) dimensional strongly coupled field theories. There it was shown that the holographic DC conductivity obeys a lower bound of the form $e^2 \sigma \geq 1$ where $e^2$ is the bulk Maxwell coupling.
Therefore holographic disorder driven metal insulator transitions are absent. A different picture arises if additional couplings
are included that lead to effective renormalization of the bulk gauge coupling constant. In \cite{Baggioli:2016oqk,Gouteraux:2016wxj} it was found that couplings
between the massless scalar fields and the Maxwell field lead to conductivities that are unbounded from below, even becoming
negative for certain ranges of the parameters and thus indicating some instability of the system. 

Here we will be interested to generalize the setup of \cite{Baggioli:2016oqk,Gouteraux:2016wxj} to strongly coupled theories in (3+1) dimensions described by
holography. There are a couple of differences to the (2+1) dimensional setup. First of all the scaling dimension of the conductivity
changes. More importantly (3+1) dimensional field theories with chiral fermions have chiral anomalies which in holography are
represented by five dimensional Chern-Simons terms. Two such terms are relevant: a pure gauge Chern-Simons term for the usual
chiral anomaly and a mixed gauge gravitational Chern-Simons term for the gravitational contribution to the anomaly. 
Inclusion of these terms is known to give rise to the chiral magnetic and chiral vortical effects (CME and CVE). 
These currents are a direct consequence of the anomalies and thus are protected from quantum corrections and give rise
to dissipationless transport.
One would therefore guess that their values have to be independent of the graviton mass and even the charged disorder parameter
introduced in \cite{Baggioli:2016oqk,Gouteraux:2016wxj}. The purpose of this work is to check this in a simple explicit example, and indeed we find that the
CME and CVE are completely insensitive to the holographic disorder parameters. In order to arrive at this result we need however 
to re-analyze the form of the holographic energy-momentum tensor. For the particular case of the massless scalar field background
under consideration the usual expression of the holographic energy-momentum tensor receives a correction due to the presence of 
the mixed gauge gravitational Chern-Simons term. Only by using this corrected form of the energy-momentum tensor we find the
expected result that the anomalous transport is insensitive to the holographic disorder. This solves the puzzle found in \cite{Megias:2016aje} based on the uncorrected form of the energy-momentum tensor.

This work is organized as follows: we introduce the model and fix conventions in section~\ref{sec:model}. In section~\ref{sec:cmecve} we calculate the 
electric DC and the chiral magnetic and chiral vortical conductivities. We conclude and discuss our results in section~\ref{sec:discussion}. 

\section{Holographic momentum relaxation}\label{sec:model}
We will base our considerations on the St\"uckelberg mechanism using massless scalar fields (Goldstone modes of
translation symmetry breaking). For each spatial dimension we introduce a scalar field $X^I$, $I=1,\dots, 3$. 
A spatially linear profile of these scalars provides the mass for the gravitons. The action of our model is
\begin{equation}
S = \int d^5x \sqrt{-g} \left[
R + 12 - \frac 1 2 \partial^\mu X^I\partial_\mu X^I - \frac 1 4 F^2  - \frac J 4 \partial_\mu X^I \partial_\nu X^I F^\mu\,_\lambda F^{\lambda\nu} 
\right] + S_\mathrm{GH} + S_\mathrm{CS} + S_\mathrm{CSK}\,.
\end{equation}
For simplicity we have set the scale of the AdS radius, the gravitational coupling $\frac{1}{2\kappa^2}$ and the bulk gauge
coupling $e$ to one. The latter can easily be re-introduced by scaling the gauge field accordingly. 
The coupling $J$ represents the effects of disorder on the charged sector of the theory \cite{Baggioli:2016oqk,Gouteraux:2016wxj}. Our model is a particular
simple case of the class of models studied there. $S_\mathrm{GH}$ is the usual Gibbons-Hawking counterterm. 
In addition we introduce a Chern-Simons action and counterterm
\begin{align}
S_\mathrm{GH} &= 2 \int_\partial d^4x \sqrt{\gamma} K\,,\\
S_\mathrm{CS} &= \int d^5x \sqrt{-g} \epsilon^{\mu\nu\rho\lambda\sigma} A_\mu \left( 
\frac{\alpha}{3} F_{\nu\rho}F_{\lambda\sigma} + \lambda R^\alpha\,_{\beta\mu\nu} R^\beta\,_{\alpha\lambda\sigma}
\right)\,,\\
S_\mathrm{CSK} &= -8 \lambda \int_\partial d^4x \sqrt{\gamma} n_\mu \epsilon^{\mu\nu\rho\lambda\sigma} A_\nu K_{\rho\tau} D_\lambda K^\tau_\sigma\,,
\end{align}
where $\gamma_{\mu\nu}$ is the induced metric and $K_{\mu\nu}$ is the extrinsic curvature on the boundary defined by an outward pointing unit normal vector. 
The gauge variation of the Chern-Simons action is a total derivative and leads to the anomaly 
\begin{equation}
\delta S = \int_\partial d^4x \sqrt{-\gamma} \epsilon^{ijkl} \left( 
\frac{\alpha}{3}  F_{ij}F_{kl}  + \lambda R^a\,_{bij} R^b\,_{akl}
\right)\,.
\end{equation}
Here $R^a\,_{bij}$ is the intrinsic curvature on the boundary. This form holds for finite holographic cutoff only if the counterterm $S_\mathrm{CSK}$
is added to the action \cite{Landsteiner:2011iq}.

The equations of motion are
\begin{align}
0=& G_{\mu\nu}-6 g_{\mu\nu} + \frac{1}{2} F_{\mu}\,^{\lambda}F_{\lambda\nu} - \frac{1}{8} g_{\mu\nu} F^2 - \frac{1}{2} \partial_\mu X^I\partial_\nu X^I +
\frac 1 4 g_{\mu\nu} (\partial_\rho X^I)(\partial^\rho X^I) - \\
& \frac J 4 \left(\tilde X.F.F+F.\tilde X.F+ F.F\tilde X \right)_{\mu\nu} 
+ \frac J 8 g_{\mu\nu} \mathrm{tr}(\tilde X.F.F) - 2 \lambda \epsilon_{\alpha\beta\gamma\delta(\mu} \nabla_\rho \left(
 F^{\beta\alpha}R^\rho\,_{\nu)}\,^{\gamma\delta}\right)\,, \nonumber \\
0=& \nabla_\mu F^{\mu\nu} + \frac J 2 \nabla_\lambda(\tilde X . F)^{\nu\lambda} - \frac J 2 \nabla_\lambda(\tilde X. F)^{\lambda\nu} + \epsilon^{\nu\alpha\beta\gamma\delta} \left( \alpha F_{\alpha\beta} F_{\gamma\delta} + \lambda R^{\rho}\;_{\sigma\alpha\beta} R^{\sigma}\,_{\rho\gamma\delta} \right) \,,\\
0=& \Box X^I + \frac{J}{2} \nabla_\mu ( \partial_\nu X^I F^\nu\,_\lambda F^{\lambda\mu})\,.
\end{align}
Following \cite{Gouteraux:2016wxj} we have defined $\tilde X_{\mu\nu}  = \partial_\mu X^I \partial_\nu X^I$ such that we can multiply 
$\tilde X.F.F =  \partial_\mu X^I \partial_\nu X^I F^\nu\,_\lambda F^{\lambda\mu} $, etc. Momentum dissipation is implemented by giving the scalar fields a linear profile (note that there are only derivative couplings of $X^I$).
We use three scalars for the three spatial dimensions and choose
\begin{equation}
 X^1 = k x ~,~~~~~~~~ X^2 = k y ~,~~~~~~~~ X^3 = k z \,. 
\end{equation}
Because the scalars couple only through derivatives, the field
equations and solutions will still be formally translational
invariant. The parameter $k$ gives the graviton however a mass and
this suffices to make the DC conductivity of a charged black hole
solution finite. As we will see, it also fixes a preferred frame (the
rest frame of the impurity density) and no frame ambiguity in the
definition of the anomalous transport coefficients appears
\cite{Amado:2011zx, Rajagopal:2015roa,Stephanov:2015roa}.

In order to define the background we look for charged black hole solutions with AdS asymptotics of the form
\begin{align}
ds^2 &= \frac{1}{u}\left(-f(u) dt^2 + dx^2+dy^2+dz^2\right) + \frac{du^2}{4u^2f(u)}\,,\\
A_t &= \phi(u)\,. 
\end{align}
The equations of motion for the blackening factor and the time component of the gauge field are
\begin{align}
0&=  f'(u)  + \frac{2}{u} (1-f(u)) - \frac{u^2}{3} ( \phi'(u))^2 - \frac{k^2}{4} \,, \\
0&= \phi''(u)\,.
\end{align}
We fix two integration constants by demanding that $\phi$ vanishes at the horizon and that the horizon sits at $u=1$.
The solutions are then
\begin{align}
f(u) &= (1-u) \left(1+u -\frac{k^2}{4} u - \frac{\mu^2}{3} u^2 \right) \,, \\
\Phi(u) &= \mu(1-u)\,.
\end{align}
We identify $\mu$ with the chemical potential. The temperature is
\begin{equation}\label{eq:T}
T = \frac{1}{\pi} \left( 1 - \frac{k^2}{8} - \frac{\mu^2}{6} \right)\,.
\end{equation}
In this simple theory the background does not depend on the charge disorder coupling $J$. It therefore reduces to the solution
found in \cite{Andrade:2013gsa}. 

\subsection{Holographic energy-momentum tensor}
The mixed gravitational Chern-Simons term makes the analysis of the holographic dictionary much more complicated than
in the standard case. This is due in part to the fact that it is a higher derivative term. We want to treat it in
an effective field theory spirit, i.e. in a perturbative way and compute the leading effects in the gravitational anomaly
coupling~$\lambda$. Even to this order we need to take into account that the standard definitions of the holographic operators
might change. In particular the standard Brown-York stress tensor has to be supplemented with extra terms coming from the gravitational Chern-Simons action.
Furthermore, the higher derivative nature of the theory entails that two independent modes emerge in the asymptotic expansion of the metric near the boundary.
Following the standard dictionary these can be understood as sources for two different operators.
However, it is in principle possible to impose asymptotic boundary conditions in which one of these modes is absent and the extrinsic curvature is a functional of the near-boundary metric. The variation of the on-shell action with respect to the boundary metric gives then the stress tensor of the dual theory.

Let us outline the main steps in the construction of the stress tensor in our case. For convenience we switch in this section
to the standard Fefferman-Graham coordinates
\begin{equation}
ds^2 = dr^2 + \gamma_{ij} dx^i dx^j
\end{equation}
where the asymptotic AdS expansion reads
\begin{equation}\label{eq:asymp}
\gamma_{ij} = e^{2r} \gamma^{(0)}_{ij} + \gamma^{(-2)}_{ij} + e^{-2r} \gamma^{(-4)}_{ij} + \dots\,.
\end{equation} 
Choosing a timelike hypersurface at a fixed $r$, we vary the on-shell action: 
\begin{equation}
\delta S = \frac{1}{2}\int_\partial \sqrt{-\gamma} \left( t^{ij} \delta \gamma_{ij} + u^{ij}\delta K_{ij} \right) + \delta S_{matter}\, .
\end{equation}
Notice that for now we keep $\gamma_{ij}$ and $K_{ij}$ as independent variable. Only at the end of the construction we will restrict ourselves to the asymptotic expansion \eqref{eq:asymp}.
We will rephrase the bulk constraint equations as the Ward identities of the dual theory. This allows us to find a bulk expression for the stress tensor by comparison with the QFT expression. 
As we expect the theory to contain a further operator on general solutions, we will allow the extrinsic curvature to only enter, outside of the stress tensor, as a source term. Consistency however demands such contribution to vanish asymptotically once \eqref{eq:asymp} is considered.

Using thus the standard forms of the variations $\delta \gamma_{ij}$ and $\delta K_{ij}$ under diffeomorphisms one finds the Ward identity:
\begin{equation}\label{eq:Wardid}
D_i\left( t^i\,_j + u^{il} K_{lj} \right) = u^{il}D_j K_{il} + F^{ij} J_i -Y^I \partial_j X_I - 8\lambda D_k \left( \epsilon^{lmnp} F_{lm} {R^{k}}_{jnp}\right) \, , 
\end{equation}
where 
\begin{equation}
Y^I= \sqrt{-\gamma}\left[\dot{X}^I + \frac{J}{2}\left( F^{ri} \dot{X}^I + F^{ji} \partial_j X^I \right) F_{ir}\right] 
\end{equation}
is the momentum conjugate to $X^I$ and
\begin{equation}
J^i=\sqrt{-\gamma}\left[ F^{ir} -8\lambda \epsilon^{ijkl} K_{j}^m D_k K_{lm} + \frac{J}{2} \left( F^{ir} \partial_i X^I \partial^j X_I - \dot{X}_I \dot{X}^I F^{rj} - \dot{X}_I \partial_i X^I F^{ij} \right)\right]
\end{equation}
is the current. The latter fulfills the Ward identity
\begin{equation}\label{eq:wicur}
D_i J^i = - \epsilon^{ijkl}\left( \alpha \sqrt{-\gamma}  F_{ij} F_{kl} + \lambda  R^a\,_{bij} R^b\,_{akl} \right)\, ,
\end{equation}
To compare to a weakly coupled theory with $N_\chi$ chiral fermions we can identify
\begin{equation}\label{eq:compare}
\alpha = \frac{N_\chi}{32\pi^2} ~~~,~~~ \lambda = \frac{N_\chi}{768\pi^2}\,.
\end{equation}
Notice that this definitions contain terms proportional to the couplings $J$ and $\lambda$ which vanish at the conformal boundary. 
Due to the fact that it obeys the correct form of Ward identity we define the holographic energy-momentum tensor as
\begin{equation}\label{eq:emtensornew}
{\Theta^i}_j =  t^i\,_j + u^{il} K_{lj} \,.
\end{equation}
We also note that $\Theta^{ij}$ and $J^i$ are the covariant currents in the sense of \cite{Bardeen:1984pm} and 
(\ref{eq:Wardid}), (\ref{eq:wicur}) are the covariant anomalies.

The tensor $t^{ij}$ can furthermore be divided into the standard Brown-York contribution $t^{ij}_0$ and a part that stems from the 
mixed gauge gravitational Chern-Simons term $t^{ij}_\lambda$. A lengthy but straightforward calculation gives
\begin{align}
 t_0^{ij} &=  - 2 \sqrt{-\gamma} ( K^{ij} - K \gamma^{ij} ) \, \\
 t_\lambda^{ij} &=- 8\lambda \sqrt{-\gamma} \epsilon^{mnp(i} \left( 2 D_n K_p^{j)} F_{rm} + \gamma^{j)l} \dot{K}_{ln} F_{pm} - F_{pm} K_l^{j)} K_n^l \right) \, , \\
 u^{ij} &= 8 \lambda \sqrt{-\gamma} \epsilon^{mnp(i} F_{mn} K_p^{j)} \, , 
\end{align}
where round brackets indicate symmetrization $A_{(ij)} = \frac{1}{2}\left(A_{ij} + A_{ji}\right)$.
 
The additional contributions will be essential to get the physically correct results for the anomalous transport coefficients.  
This generalizes the expression found in \cite{Megias:2013joa} to cases in which $\gamma^{(-2)}$ does not vanish. 
We also note that $(t_0)_j^i$ is divergent and needs to be regularized by the standard counterterms \cite{deHaro:2000vlm}. 
In contrast, the additional contributions $(t_\lambda)^i_j$ and $u^{il}K_{lj}$ are already finite before the holographic renormalization is performed.

In principle the correction can also be found by noting that the variation of $K_{ij}$ under the assumption of the
asymptotic expansion (\ref{eq:asymp}) is not independent from the variation of $\gamma_{ij}$.
We note that in the case of holographic pure gravitational anomalies dual to two dimensional field theories a similar correction
has been found in \cite{Kraus:2005zm}. 

This definition of the stress tensor, together with the further contribution to the Ward identity \eqref{eq:Wardid} are nontrivial consequences of the dynamics of the anomalous theory. An in depth analysis of their properties will be given 
in a forthcoming publication.

\section{Electric and chiral magnetic conductivities}\label{sec:cmecve}
In this section we will compute the electric DC conductivity and the chiral magnetic and chiral vortical
conductivities in linear response theory. We start with the
\subsection{DC conductivity}
We introduce the small perturbations
\begin{align}
A_z &= \epsilon (- E t + a_z(u) )\,,\\
g_{tz} &= \frac{\epsilon}{u}  h_t^z(u) \,,\\
g_{uz} &= \frac{\epsilon}{u}  h_u^z(u) \,.  
\end{align}
$E$ is the external electric field and the perturbations all fulfill $a_z(0)=h_t^z(0)=h_u^z(0)=0$, i.e. they do not
introduce additional sources. 
The equations of motion are
\begin{align}
0&=\left(1+2 u^3 J \phi'(u)^2 \right) h^z_u + E \frac{u \phi'(u)}{k^2 f(u)} \left( 1 - \frac{k^2}{2} J u\right)\,,\\
u^2 f(u) \left( \frac{h'^z_t(u)}{u}\right)' &= \frac{k^2}{4}\left(1+2u^3 J \phi'(u)^2\right) h_t^z -  u^2\phi'(u)\left(
1-\frac{k^2}{2}J u\right) f(u) a_z'(u)\,,\\
0&=\left[\left(1-\frac{k^2}{2} J u\right)\left(f(u)a_z'(u) +\phi'(u) h_t^z(u)\right) \right]' \,.
\end{align}
We solve these equations following \cite{Donos:2014uba}. Note that $h^z_u(u)$ obeys an algebraic equation. Demanding regularity
of the metric we find that as one approaches the horizon  
\begin{equation}
2 f(u) h_u^z(u)  = - h_t^z(u)\,.
\end{equation}
The equation for $a_z(u)$ is a total derivative and can be integrated. On the horizon we can compute $f(u) a_z'(u)$ using the
equation for $h^z_t$. Noting that the current is $J_z = 2 \lim_{u\rightarrow 0} f(u) a_z'(u)$ we obtain the 
conductivity $\sigma_\mathrm{DC} = J/E$  
\begin{equation}\label{eq:DCcond}
\sigma_\mathrm{DC} = \left(1-\frac{k^2}{2} J \right) \left[ 
1 + \frac{\left(1-\frac{k^2}{2} J\right) 4 \mu^2}{k^2\left(1+2 J \mu^2 \right)}
\right] \,.
\end{equation}
This expression is the dimensionless conductivity where we have set the horizon to $u_h=1$. It is qualitatively 
of the same form as the one discussed in the $AdS_4$ model in \cite{Gouteraux:2016wxj}, e.g. it vanishes 
for $k^2 J =2$ and can become
even negative in some range of the parameters indicating an instability. We also note that for $J=0$ the dimensionless 
conductivity obeys a similar bound as the one proven for holographic matter in 2+1 dimensions in \cite{Grozdanov:2015qia}.
Our main interest is however in the anomalous transport
coefficients so we will not further investigate the properties of $\sigma_{DC}$. 

\subsection{Chiral magnetic conductivity}
We now introduce the magnetic field as a perturbation by
\begin{align}
A_y &= \epsilon B x \,,\\
A_z &= \epsilon a_z(u)\,,\\
g_{tz} &=  \frac{\epsilon}{u} h_t^z(u) \,.  
\end{align}
The equations of motion for the perturbations are
\begin{align}
4 \alpha B  \phi'(u) =& \left[ \left(1-\frac{k^2}{2} J u \right) \left( f(u) a_z'(u) + \phi'(u) h^t_z(u) \right) \right]'\,, \\
u^2 f(u) \left( \frac{h'^z_t(u)}{u}\right)' =& \frac{k^2}{4}\left(1+2u^3 J \phi'(u)^2\right) h_t^z -  u^2\phi'(u)\left(
1-\frac{k^2}{2}J u\right) f(u) a_z'(u) \nonumber\\ 
 &-2 B \lambda f(u) \left(3 u k^2+16 u^2  \phi'(u)^2 +12 f(u) -  12  \right) \,.
\end{align}
The strategy to integrate these equations is as follows. First we solve the equations for $a_z(u)$ by writing
\begin{equation}
h^t_z(u) = \frac{f(u)}{\mu} a_z'(u) + \frac{4 \alpha B (u-1)}{1-  \frac{k^2}{2}J u} \,,
\end{equation}
where we have fixed an integration constant by demanding that $h^t_z(u)$ vanishes on the horizon. The equation for $h^t_z(u)$ is
now converted into a third order equation for $a_z(u)$. This equation can be integrated and the remaining three integration constants
can be fixed by demanding regularity of the solutions on the horizon. Due to the presence of the mass terms for the graviton fluctuation, this means in particular that $h^t_z(u)$ has to vanish on the horizon. We note that for $k^2=0$ this
is not the case and gives rise to an additional integration constant that eventually corresponds to the choice of frame in a hydrodynamic setup. Without going into the details of the solution, we note that this procedure results in the asymptotic
expansions
\begin{align}
a_z(u) &= 4 \alpha B \mu u + O(u^2) \,,\\
h^z_t(u) &= -\left(\alpha \mu^2  + 8 \lambda \pi^2 T^2  - \lambda \frac{k^2}{2}\right) B u^2 + O(u^3)\,.
\end{align} 
The result is almost independent of the disorder parameters $k$ and $J$. However if we apply the usual holographic dictionary
and compute the energy-momentum tensor from these solutions we would find $T_{0z} = 4 g_{tz}'(u=0)$ \cite{deHaro:2000vlm}. 
As we have argued in the previous section this corresponds only to the contribution $t^{ij}_0$, and we need also to calculate the corrections due to $(t_\lambda)^{ij}$ and $u^{il}K_{lj}$. These can be easily calculated from the
background and give
\begin{align}
(t_\lambda)^{ij} &=0\,,\\
(u.K)^{ij} &= 2 k^2 B \lambda \delta^{i(0}\delta^{z)j}  \,.
\end{align} 
Using therefore the new definition of energy-momentum tensor (\ref{eq:emtensornew}) and assembling the components $\Theta^{i0}$ into the energy current $\vec J_E$,  we find
\begin{align}
\vec J &= 8 \alpha \mu \vec B\,,\\
\vec J_E &= (4 \alpha \mu^2 + 32\lambda  \pi^2 T^2) \vec B\,.
\end{align}
Taking into account (\ref{eq:compare}) these are the usual expressions for the chiral magnetic effect in (covariant) charge and 
currents.

\subsection{Chiral vortical conductivity}
We will represent vorticity by introducing a gravitomagnetic field $B_g$ in the $z$ direction.
The relation between vorticity and the gravitomagnetic field follows from observing that $\Omega^i = \frac{1}{2} \epsilon^{ijk} \partial_j u_k$. In the rest frame $u^\mu = (1,0,0,0)$, the gravitomagnetic field is the mixed space-time component of the metric in the dual field theory $ds^2 = -dt^2 + 2 dt \vec{A}_g d\vec x + d\vec x^2 $. With the gravitomagnetic field defined as $\vec{B}_g = \vec \nabla \times \vec{A}_g$, it follows $\vec B_g = 2\vec \Omega$. The ansatz for the perturbations is
\begin{align}
A_y &= \epsilon B_g u \mu x \,,\\
A_z &= \epsilon a_z(u)\,,\\
g_{ty} &= \epsilon \frac{f(u)}{u} B_g x \,,\\
g_{tz} &=  \epsilon \frac{1}{u} h_t^z(u) \,.  
\end{align}
We note that the gravitomagnetic field produces a current due to drag when the chemical potential does not vanish. The equations of motion  for the perturbations are
\begin{align}
4 \alpha B_g u \phi'(u)^2 + B_g \lambda & \left(20 f'(u)^2 - 16 (f(u)-1)\frac{f'(u)}{u} + \frac{16}{3}u^2 f'(u) \phi'(u)^2\right)  
\nonumber \\
= &\left[ \left(1-\frac{k^2}{2} J u \right) \left( f(u) a_z'(u) + \phi'(u) h^t_z(u) \right) \right]' \,,
  \\
u^2 f(u) \left( \frac{h'^z_t(u)}{u}\right)' = &\frac{k^2}{4}\left(1+2u^3 J \phi'(u)^2\right) h_t^z -  u^2\phi'(u)\left(
1-\frac{k^2}{2}J u\right) f(u) a_z'(u)  \nonumber \\
&-\lambda B_g \mu  f(u) \left(  80f(u)  + 17 u  k^2+(172/3)  u^3 \mu^2 - 80    \right) \,.
\end{align}
The strategy to integrate these equations is the same as before. First we solve the equations for $a_z(u)$ by writing
\begin{align}
h^t_z(u) =& \frac{f(u)}{\mu} a_z'(u) + 32 B_g \frac{1-u}{1-\frac{k^2}{2} J u} \big(
(-16 u^2-16 u- 16 - 144 u^4 + 48 u^3) \lambda \mu^3   \nonumber \\
&+((-u-1) 72 \alpha+(576 u^3-24 u k^2-144 u^3 k^2+192 u^2-24 k^2+192 u+24 u^2 k^2+192) \lambda) \mu  \nonumber\\
& +\frac{\lambda}{\mu} (-576 u-36 u^2 k^4+144 k^2-576 u^2+288 u^2 k^2-576-9 k^4+144 u k^2)
\big) \,.
\end{align}
Again we have fixed an integration constant by demanding that $h^t_z(u)$ vanishes on the horizon. The equation for $h^t_z(u)$ is
now converted into a third order equation for $a_z(u)$. This equation can be solved with the same boundary conditions as
in the case of the magnetic field. The details of the lengthy solutions are not interesting to us. The resulting asymptotic expansions are
\begin{align}
a_z(u) &= ( 2 \alpha \mu^2 u + 16 \pi^2 T^2 \lambda) B_g u + O(u^2) \,,\\
h^z_t(u) &= -\left(\frac{2}{3} \alpha \mu^3  + 16 \lambda \mu \pi^2 T^2 \right) B_g u^2 + O(u^3)\,.
\end{align} 
In contrast to the case with magnetic field, the asymptotic expansions are completely independent of the disorder parameters. 
This is accord with the form of the energy-momentum tensor since the new contributions to it depend only on external electromagnetic fields but not
on the externally applied gravitomagnetic field. 
We find thus for the response due to a gravitomagnetic field
\begin{align}
\vec J &= \left( 4 \alpha \mu^2 + 32 \pi^2 T^2 \lambda \right)  \vec B_g \,,\\
\vec J_E &= \left(\frac{8}{3} \alpha \mu^3 + 64\lambda  \mu\pi^2 T^2 \right) \vec B_g\,.
\end{align}
Remembering that gravitomagnetic field and vorticity are related by $\vec B_g = 2 \vec\Omega$, these are the usual responses of a chiral fluid due to vorticity. 
 
\section{Conclusion and Discussion}\label{sec:discussion}

Chiral magnetic and chiral vortical effects are anomaly induced dissipationless transport phenomena. Holography has played a major role in their discovery and interpretation. Here we showed that - as expected - in a simple holography model
of disorder the chiral magnetic and chiral vortical effects are unchanged. 

While this result could have been expected and at least the response in the charge current could be inferred from~\cite{Grozdanov:2016ala}, the response in the energy current turned out to be more interesting. The mixed gauge gravitational
Chern-Simons term modifies the definition of the holographic energy-momentum tensor. We found the new form by allowing the extrinsic curvature to vary independently near the boundary. Then we demanding that the energy momentum tensor fulfills
a standard Ward identity in which the extrinsic curvature acts like an external source conjugate to the operator $u^{ij}$. We emphasize however that the background we found is a perfectly asymptotically AdS background in which the extrinsic curvature is asymptotically proportional to the induced metric in the standard fashion $K_{ij} = \gamma_{ij} + \dots$. Nevertheless, even to first order in the gravitational Chern-Simons coupling~$\lambda$, the new terms in the energy-momentum tensor contribute and precisely restore the usual form of the chiral magnetic effect in the energy current. It is also important to realize that the new terms contribute only due to the fact that the asymptotic expansion of the metric contains a constant term, $g_{tt} = -\frac 1 u + \frac{k^2}{4} + O(u)$ and $g_{uu} = \frac{1}{4u} + \frac{k^2}{16} + O(u)$, which is a direct consequence of the presence of a background of massless scalar fields. 

Another important remark concerns the form of the anomalous transport coefficients. In translation symmetry preserving backgrounds, an additional integration constant in the solution for the fluctuation $h_t^z$ is allowed. This integration constant can then be used to choose a frame, e.g. one could use it to fix the Landau frame \cite{Son:2009tf}. In the theory with a massive graviton, such an integration constant is absent and regularity of the solution uniquely fixes the frame which should be interpreted as the disorder rest frame \cite{Rajagopal:2015roa,Stephanov:2015roa}. 

We also note that the gravitational Chern-Simons term and the charge
disorder term proportional to the coupling $J$ are both of higher
derivative nature and therefore should in principle be understood as
perturbative couplings in an effective field theory manner, viewing
them as subleading terms in the $1/N$ expansion.  Since we are
only interested in linear response in the magnetic and
gravito-magnetic fields, this is automatic for the Chern-Simons
coupling~$\lambda$.  We decided however to follow
\cite{Baggioli:2016oqk,Gouteraux:2016wxj} who treat the analogues of
the $J$ coupling non-perturbatively. This is interesting since it
allows to compare the DC conductivity (\ref{eq:DCcond}) with the ones
obtained there.  We note that there is an unphysical region in the
parameter space in which the DC conductivity becomes negative. It is
remarkable that the anomalous transport coefficients can be computed
exactly for all values of $J$ and are indeed independent of it.

It is also interesting to recall the results of
\cite{Jimenez-Alba:2014iia} in which a St\"uckelberg field was added
to the gauge sector, thereby breaking gauge invariance in the bulk. In
this theory the chiral magnetic conductivity did get corrected.  In
contrast the St\"uckelberg field giving a mass to the graviton does
not influence the anomalous conductivities.  Intuitively we propose to
understand that as follows.  The massive gravity theory breaks
spatial translation invariance. There is however no anomalous conductivity
related to the conserved momentum. The corresponding currents would be
the purely spatial components of the energy momentum tensor.  
Threfore the anomalous terms in
$\Theta^{0i}$ should be interpreted as energy current and not
as the (non-conserved) momentum density. In our setup energy and charge are still 
conserved up to the anomalies. The anomalous transport in 
energy and charge current is not modified. In this line of reasoning it would be
very interesting to study holographic setups that implement energy
non-conservation.

Finally we note that the new form of the energy-momentum tensor should be useful to understand the effects of the mixed
gravitational Chern-Simons term on transport as has been recently studied for other higher derivative actions in \cite{Donos:2017oym}.


\section{Acknowledgments}
We would like to thank O. Pujol\`as, K. Schalm, Y. Liu, M. Baggioli and M. Valle
for useful discussions. This research has been supported by Plan Nacional de Altas Energ\'{\i}as Spanish MINECO grants FPA2015-65480-P and FPA2015-64041-C2-1-P, by the Basque Government grant IT979-16, and by the Centro de Excelencia Severo Ochoa Programme under grant SEV-2012-0249. The work of J.F.-P. is supported by
fellowship SEV-2012-0249-03. The work of C.C. is funded by Fundaci\'on
La Caixa under ``La Caixa-Severo Ochoa'' international predoctoral
grant. The work of E.M. is supported by the Universidad del Pa\'{\i}s Vasco UPV/EHU, Bilbao, Spain, as a Visiting Professor.

\bibliography{AnomTrans}{}

\providecommand{\href}[2]{#2}\begingroup\raggedright\begin{thebibliography}{10}

\bibitem{Kharzeev:2013ffa}
D.~E. Kharzeev, \emph{{The Chiral Magnetic Effect and Anomaly-Induced
  Transport}}, \href{http://dx.doi.org/10.1016/j.ppnp.2014.01.002}{\emph{Prog.
  Part. Nucl. Phys.} {\bf 75} (2014) 133--151},
  [\href{https://arxiv.org/abs/1312.3348}{{\tt 1312.3348}}].

\bibitem{Landsteiner:2016led}
K.~Landsteiner, \emph{{Notes on Anomaly Induced Transport}},  in \emph{{56th
  Cracow School of Theoretical Physics: A Panorama of Holography Zakopane,
  Poland, May 24-June 1, 2016}}, 2016.
\newblock \href{https://arxiv.org/abs/1610.04413}{{\tt 1610.04413}}.

\bibitem{Newman:2005hd}
G.~M. Newman, \emph{{Anomalous hydrodynamics}},
  \href{http://dx.doi.org/10.1088/1126-6708/2006/01/158}{\emph{JHEP} {\bf 01}
  (2006) 158}, [\href{https://arxiv.org/abs/hep-ph/0511236}{{\tt
  hep-ph/0511236}}].

\bibitem{Banerjee:2008th}
N.~Banerjee, J.~Bhattacharya, S.~Bhattacharyya, S.~Dutta, R.~Loganayagam and
  P.~Surowka, \emph{{Hydrodynamics from charged black branes}},
  \href{http://dx.doi.org/10.1007/JHEP01(2011)094}{\emph{JHEP} {\bf 01} (2011)
  094}, [\href{https://arxiv.org/abs/0809.2596}{{\tt 0809.2596}}].

\bibitem{Erdmenger:2008rm}
J.~Erdmenger, M.~Haack, M.~Kaminski and A.~Yarom, \emph{{Fluid dynamics of
  R-charged black holes}},
  \href{http://dx.doi.org/10.1088/1126-6708/2009/01/055}{\emph{JHEP} {\bf 01}
  (2009) 055}, [\href{https://arxiv.org/abs/0809.2488}{{\tt 0809.2488}}].

\bibitem{Landsteiner:2011iq}
K.~Landsteiner, E.~Megias, L.~Melgar and F.~Pena-Benitez, \emph{{Holographic
  Gravitational Anomaly and Chiral Vortical Effect}},
  \href{http://dx.doi.org/10.1007/JHEP09(2011)121}{\emph{JHEP} {\bf 09} (2011)
  121}, [\href{https://arxiv.org/abs/1107.0368}{{\tt 1107.0368}}].

\bibitem{Ammon:2015wua}
M.~Ammon and J.~Erdmenger, \emph{{Gauge/gravity duality}}.
\newblock Cambridge Univ. Pr., Cambridge, UK, 2015.

\bibitem{Zaanen:2015oix}
J.~Zaanen, Y.-W. Sun, Y.~Liu and K.~Schalm, \emph{{Holographic Duality in
  Condensed Matter Physics}}.
\newblock Cambridge Univ. Press, 2015.

\bibitem{Vegh:2013sk}
D.~Vegh, \emph{{Holography without translational symmetry}},
  \href{https://arxiv.org/abs/1301.0537}{{\tt 1301.0537}}.

\bibitem{Davison:2013jba}
R.~A. Davison, \emph{{Momentum relaxation in holographic massive gravity}},
  \href{http://dx.doi.org/10.1103/PhysRevD.88.086003}{\emph{Phys. Rev.} {\bf
  D88} (2013) 086003}, [\href{https://arxiv.org/abs/1306.5792}{{\tt
  1306.5792}}].

\bibitem{Blake:2013bqa}
M.~Blake and D.~Tong, \emph{{Universal Resistivity from Holographic Massive
  Gravity}}, \href{http://dx.doi.org/10.1103/PhysRevD.88.106004}{\emph{Phys.
  Rev.} {\bf D88} (2013) 106004}, [\href{https://arxiv.org/abs/1308.4970}{{\tt
  1308.4970}}].

\bibitem{Blake:2013owa}
M.~Blake, D.~Tong and D.~Vegh, \emph{{Holographic Lattices Give the Graviton an
  Effective Mass}},
  \href{http://dx.doi.org/10.1103/PhysRevLett.112.071602}{\emph{Phys. Rev.
  Lett.} {\bf 112} (2014) 071602}, [\href{https://arxiv.org/abs/1310.3832}{{\tt
  1310.3832}}].

\bibitem{Donos:2013eha}
A.~Donos and J.~P. Gauntlett, \emph{{Holographic Q-lattices}},
  \href{http://dx.doi.org/10.1007/JHEP04(2014)040}{\emph{JHEP} {\bf 04} (2014)
  040}, [\href{https://arxiv.org/abs/1311.3292}{{\tt 1311.3292}}].

\bibitem{Andrade:2013gsa}
T.~Andrade and B.~Withers, \emph{{A simple holographic model of momentum
  relaxation}}, \href{http://dx.doi.org/10.1007/JHEP05(2014)101}{\emph{JHEP}
  {\bf 05} (2014) 101}, [\href{https://arxiv.org/abs/1311.5157}{{\tt
  1311.5157}}].

\bibitem{Grozdanov:2015qia}
S.~Grozdanov, A.~Lucas, S.~Sachdev and K.~Schalm, \emph{{Absence of
  disorder-driven metal-insulator transitions in simple holographic models}},
  \href{http://dx.doi.org/10.1103/PhysRevLett.115.221601}{\emph{Phys. Rev.
  Lett.} {\bf 115} (2015) 221601},
  [\href{https://arxiv.org/abs/1507.00003}{{\tt 1507.00003}}].

\bibitem{Baggioli:2016oqk}
M.~Baggioli and O.~Pujolas, \emph{{On holographic disorder-driven
  metal-insulator transitions}},
  \href{http://dx.doi.org/10.1007/JHEP01(2017)040}{\emph{JHEP} {\bf 01} (2017)
  040}, [\href{https://arxiv.org/abs/1601.07897}{{\tt 1601.07897}}].

\bibitem{Gouteraux:2016wxj}
B.~Goutéraux, E.~Kiritsis and W.-J. Li, \emph{{Effective holographic theories
  of momentum relaxation and violation of conductivity bound}},
  \href{http://dx.doi.org/10.1007/JHEP04(2016)122}{\emph{JHEP} {\bf 04} (2016)
  122}, [\href{https://arxiv.org/abs/1602.01067}{{\tt 1602.01067}}].

\bibitem{Megias:2016aje}
E.~Megias, \emph{{Anomalous transport and massive gravity theories}},
  \href{http://dx.doi.org/10.1088/1742-6596/804/1/012029}{\emph{J. Phys. Conf.
  Ser.} {\bf 804} (2017) 012029}, [\href{https://arxiv.org/abs/1701.00085}{{\tt
  1701.00085}}].

\bibitem{Amado:2011zx}
I.~Amado, K.~Landsteiner and F.~Pena-Benitez, \emph{{Anomalous transport
  coefficients from Kubo formulas in Holography}},
  \href{http://dx.doi.org/10.1007/JHEP05(2011)081}{\emph{JHEP} {\bf 05} (2011)
  081}, [\href{https://arxiv.org/abs/1102.4577}{{\tt 1102.4577}}].

\bibitem{Rajagopal:2015roa}
K.~Rajagopal and A.~V. Sadofyev, \emph{{Chiral drag force}},
  \href{http://dx.doi.org/10.1007/JHEP10(2015)018}{\emph{JHEP} {\bf 10} (2015)
  018}, [\href{https://arxiv.org/abs/1505.07379}{{\tt 1505.07379}}].

\bibitem{Stephanov:2015roa}
M.~A. Stephanov and H.-U. Yee, \emph{{No-Drag Frame for Anomalous Chiral
  Fluid}}, \href{http://dx.doi.org/10.1103/PhysRevLett.116.122302}{\emph{Phys.
  Rev. Lett.} {\bf 116} (2016) 122302},
  [\href{https://arxiv.org/abs/1508.02396}{{\tt 1508.02396}}].

\bibitem{Bardeen:1984pm}
W.~A. Bardeen and B.~Zumino, \emph{{Consistent and Covariant Anomalies in Gauge
  and Gravitational Theories}},
  \href{http://dx.doi.org/10.1016/0550-3213(84)90322-5}{\emph{Nucl. Phys.} {\bf
  B244} (1984) 421--453}.

\bibitem{Megias:2013joa}
E.~Megias and F.~Pena-Benitez, \emph{{Holographic Gravitational Anomaly in
  First and Second Order Hydrodynamics}},
  \href{http://dx.doi.org/10.1007/JHEP05(2013)115}{\emph{JHEP} {\bf 05} (2013)
  115}, [\href{https://arxiv.org/abs/1304.5529}{{\tt 1304.5529}}].

\bibitem{deHaro:2000vlm}
S.~de~Haro, S.~N. Solodukhin and K.~Skenderis, \emph{{Holographic
  reconstruction of space-time and renormalization in the AdS / CFT
  correspondence}},
  \href{http://dx.doi.org/10.1007/s002200100381}{\emph{Commun. Math. Phys.}
  {\bf 217} (2001) 595--622}, [\href{https://arxiv.org/abs/hep-th/0002230}{{\tt
  hep-th/0002230}}].

\bibitem{Kraus:2005zm}
P.~Kraus and F.~Larsen, \emph{{Holographic gravitational anomalies}},
  \href{http://dx.doi.org/10.1088/1126-6708/2006/01/022}{\emph{JHEP} {\bf 01}
  (2006) 022}, [\href{https://arxiv.org/abs/hep-th/0508218}{{\tt
  hep-th/0508218}}].

\bibitem{Donos:2014uba}
A.~Donos and J.~P. Gauntlett, \emph{{Novel metals and insulators from
  holography}}, \href{http://dx.doi.org/10.1007/JHEP06(2014)007}{\emph{JHEP}
  {\bf 06} (2014) 007}, [\href{https://arxiv.org/abs/1401.5077}{{\tt
  1401.5077}}].

\bibitem{Grozdanov:2016ala}
S.~Grozdanov and N.~Poovuttikul, \emph{{Universality of anomalous
  conductivities in theories with higher-derivative holographic duals}},
  \href{http://dx.doi.org/10.1007/JHEP09(2016)046}{\emph{JHEP} {\bf 09} (2016)
  046}, [\href{https://arxiv.org/abs/1603.08770}{{\tt 1603.08770}}].

\bibitem{Son:2009tf}
D.~T. Son and P.~Surowka, \emph{{Hydrodynamics with Triangle Anomalies}},
  \href{http://dx.doi.org/10.1103/PhysRevLett.103.191601}{\emph{Phys. Rev.
  Lett.} {\bf 103} (2009) 191601}, [\href{https://arxiv.org/abs/0906.5044}{{\tt
  0906.5044}}].

\bibitem{Jimenez-Alba:2014iia}
A.~Jimenez-Alba, K.~Landsteiner and L.~Melgar, \emph{{Anomalous magnetoresponse
  and the St{\"u}ckelberg axion in holography}},
  \href{http://dx.doi.org/10.1103/PhysRevD.90.126004}{\emph{Phys. Rev.} {\bf
  D90} (2014) 126004}, [\href{https://arxiv.org/abs/1407.8162}{{\tt
  1407.8162}}].

\bibitem{Donos:2017oym}
A.~Donos, J.~P. Gauntlett, T.~Griffin and L.~Melgar, \emph{{DC Conductivity and
  Higher Derivative Gravity}},  \href{https://arxiv.org/abs/1701.01389}{{\tt
  1701.01389}}.

\end{thebibliography}\endgroup
\bibliographystyle{JHEP}

\end{document}